# Developing an Offshore Machine Learning Surface Layer Scheme

Susan Dettling,[a] Sue Ellen Haupt,[a] Thomas Brummet,[a] Patrick Hawbecker,[a]

Branko Kosović,[b] David John Gagne[a]

[a] *NSF National Center for Atmospheric Research, Boulder, CO*

[b] *Johns Hopkins University, Baltimore, MD*

*Corresponding author*: Sue Ellen Haupt, haupt@ucar.edu

ABSTRACT

Turbulent fluxes between the surface and the atmosphere are typically parameterized using empirically fit relationships. Here we test machine learning techniques for fitting the relationship for the offshore environment. To do that, data from three offshore sites are used: the Martha's Vineyard Coastal Observatory (MVCO) air-sea interaction tower, the FINO1 research platform, and the CASPER-West FLIP research vessel deployed off the coast of California. Two machine learning methods were employed: Neural Networks (NN) and Random Forests (RF). Because the observational sites had towers with measurements at different levels, the vertical differences were input as gradients. Models were built for both momentum flux and heat flux.

ML models trained at the individual sites were competitive with and in some cases, better than the physically-based COARE-3 model tailored to offshore fluxes. The heat flux ML models generally outperformed the physics-based parameterizations for most metrics, but the results were mixed for momentum flux, with only the site with the most training data (MVCO) producing results better than COARE-3. When the ML models from that site were applied to the other sites, results were degraded from using data from the site being tested. ML models built from data combined from the three sites generally showed improvements for the sites with less available training data. When assessing which variables were most important, the wind speed was most important for momentum flux and temperature gradient for heat flux.

## 1. Introduction

Exchanges of momentum, heat, moisture, and other constituents between the surface of the Earth and the atmosphere are dominated by turbulent fluxes that cannot be resolved in numerical weather prediction (NWP) models; instead, they are parameterized using semi-empirical parameterizations. To correctly represent the atmospheric boundary layer (ABL), it is imperative to adequately represent the bottom boundary condition. The surface layer, usually taken as the lowest 10% of the ABL, is where fluxes of momentum, heat, and moisture determine the stability and resulting characteristics of the ABL and are assumed to vary less than 10% in that layer (Stull 1988). Most models represent the surface layer using Monin-Obhokov Similarity Theory (MOST- Monin & Obhukov 1954), based on Buckingham Pi similarity theory to set up a relationship between the important variable groups and then fit coefficients based on experimental data. Many of the models use coefficients based on pioneering field experiments in the grasslands of Kansas (Businger et al. 1971). Despite the fact that such grassland environments are not prevalent across the many types of environments, many models use the same coefficients globally. Additionally, MOST is based on assumptions of horizontal homogeneity and stationarity, which are seldom satisfied in nature.

Some formulations have sought to better characterize the marine environment, dating back to improved models for surface fluxes built for tropical conditions based on the TOGA-COARE experiments (Fairall et al. 1996) and later generalized to better fit subtropical environments (Fairall et al. 2003, 2011) as described in more depth below.

Given that the MOST-based formulations rely on empirical fits for the coefficients that relate the variables, the surface layer models are ripe for improvements with machine learning (ML) models of the flux processes that more directly leverage surface flux observations. Indeed, McCandless et al. (2022) used observational data to build such ML models based on both artificial neural networks (NN) and random forests (RF). They leveraged long-term observations from two different sites, a flux tower in Cabauw, The Netherlands and a second tower supported by Idaho National Laboratory. They demonstrated that both the NN and RF models were better able to capture the scales that are the basis of the flux calculations (surface friction velocity, virtual potential temperature scale, and moisture scale) than was MOST at both sites. Additionally, when the models built at one site were applied to the other, they still outperformed MOST for each of the three flux scales.

This work was further advanced by Munoz-Esparza et al. (2022) who formulated the inputs to the NN as gradients, finding that helped to generalize the results when implemented into the FastEddy large-eddy simulation model. They tested the model in idealized cases ranging from dry neutral boundary layers to a moist diurnal cycle, both offline and within FastEddy for detailed aspects of the momentum, heat and moisture flux components. The NN models improved flux predictions in comparison to MOST at the Idaho tower, correcting the systematic underestimations of turbulent fluxes, improved correlation with observations, and reduced the bias.

Note, however, that although the prior ML work on developing a ML surface layer scheme showed improvement over MOST, it still focused on flat to gently rolling grassland regions. Given the roughly 71% of the Earth covered by water, it is desirable to derive a similar ML-based surface layer model for offshore regions. Recently, Wu et al. (2026) derived a data-driven probabilistic NN air-sea flux parameterization assuming Gaussian distributions and minimizing log-likelihood loss functions and testing in a single column model.

The goal of our work is to produce an ML model of the surface layer based on flux measurements in marine environments and to evaluate the resulting model(s) according to how well they model surface fluxes as compared to MOST theory and the COARE model. Our plan is to leverage three different measurement sites and evaluate how MLSL models built for each compare and apply to the other sites. A second goal is to implement this MLSL model in the Weather Research and Forecasting (WRF - Skamarack et al. 2019) model and rigorously test its application in those several coastal environments for which we have observational data - this work is reported in the companion paper by Hawbecker et al. (2026).

In section 2 we describe our methods and data, beginning with our theoretical formulation, advancing to the data sets that were used, and describing our ML methodology. Results are presented and analyzed in section 3, comparing multiple local MLSL models and also building a ML model based on combined data from the three disparate sites. Section 4 summarizes our findings, discusses its more general applicability, lists a few caveats, and describes next steps in the research. The companion paper (Hawbecker et al. 2026) presents the results of testing our MLSL models within WRF and describes how to generalize ML approaches to improving our calculations of surface fluxes in models of the ABL.

## 2. Approach – Data and Methods

Here we describe the mathematical and computational formulations that are used in this work in section 2a. Then we present the three data sets that were used for training and how they were split for training, validation, and testing in section 2b. Section 2c treats our machine learning approaches and metrics used to assess quality of the resulting models.

### *a. Formulation*

Before going into the details of our data and machine learning methods, we set the stage for how we base our models on MOST and COARE, then provide the computation details.

1) BACKGROUND

In NWP models MOST underpins bulk parameterizations of exchange coefficients. McCandless et al. (2022) showed that over land, machine-learning-based parameterizations trained on surface observations can outperform MOST-based parameterizations. They found that over homogeneous terrain, momentum flux (or surface friction velocity) is estimated accurately with both MOST and ML models. However, McCandless et al. (2022) demonstrated that ML models perform significantly better than MOST estimating surface fluxes of moisture and sensible heat flux. MOST also represents a cornerstone of parameterizations of surface exchanges over bodies of water. However, flux parameterizations must also account for the dynamic air – sea (or lake) interactions that impact surface roughness. Charnock (1955a,b) proposed a parameterization of surface roughness as a function of surface friction velocity, $u_*$, and gravitational acceleration, g: $z_0 = \frac{\alpha u_*^2}{g}$, where α is the Charnock parameter.

Over the last three decades a bulk flux parameterization based on the Coupled Ocean-Atmosphere Response Experiment (COARE, Fairall et al. 1996) and subsequent field studies and developments (Fairall et al. 2003, 2011; Edson et al. 2013) improved on the original Charnock's parameterization. An early version of COARE parameterization (COARE 2.5) modifies stability functions and the lower limit on the roughness length corresponding to an aerodynamically rough flow, $z_0 = z_0^{rough} + z_0^{smooth} = \frac{\alpha u_*^2}{g} + 0.11\frac{\nu}{u_*}$. Here, ν is the kinematic viscosity.

The scalar (temperature and moisture) roughness lengths, $z_{0x}$, are expressed as functions of the roughness Reynolds number $R_r = \frac{z_0}{\nu}$, so that $z_{0x} = \frac{\nu_x R_r}{u_*}$ following Lui et al. (1979). The next update of the COARE algorithm (COARE 3.0) introduced a linear dependence of the parameter α on the wind speed at 10 m under neutral conditions in the range of wind speeds between 10 $ms^{-1}$ and 18 $ms^{-1}$. Below 10 $ms^{-1}$, the Charnock parameter, α=0.11, and above 18 $ms^{-1}$ it is constant. This version of the COARE algorithm includes parameterization of the scalar roughness as $z_{0q} = z_0 \exp(3.4 - 3.5 R_r^{1/4})$. A more recent update of the COARE algorithm (COARE 3.5, Edson et al. 2013) modified the linear dependence of the Charnock parameter and extended it to lower wind speeds, $\alpha = 0.017U_{10N} - 0.005$, where $U_{10N}$ is the wind speed at 10 m under neutral conditions.

Above 18 $ms^{-1}$ the Charnock parameter levels off at a value of 0.28. Edson et al. (2013) showed that the inverse wave age $\frac{u_*}{C_p}$, where $C_p$ denotes the phase speed, also varies nearly linearly with the wind speed in the range of wind speeds between 3 $ms^{-1}$ and 20 $ms^{-1}$, and therefore the commonly used dependence of the Charnock parameter on the inverse wave age, $\alpha = (A\frac{u_*}{C_p})^B$, where A and B are empirical coefficients, can be replaced with the linear dependence on the wind speed, where A and B are constants.

2) IMPLEMENTATION

We used the Drennan et al (2003) parameterization of surface roughness to allow surface roughness to dynamically adjust with sea state, $z_0 = 3.35H_s\left(\frac{u_*}{c_p}\right)^{3.4}$, where $H_s$ is wave height.

The nonlinear system of equations derived from the MOST wind and temperature profile relationships was solved using the ***fsolve*** routine from the Python SciPy library (Virtanen et al. 2020), which employs a multivariate root-finding algorithm to determine solutions given initial estimates.

An initial guess of $u_*$=0.1 m s−1 was used for the friction velocity. To ensure convergence across diverse atmospheric stability regimes, a broad set of initial guesses for the Obukhov length L was specified and evaluated sequentially until a converged solution was obtained. Convergence was defined by a relative tolerance of 10−3 applied to the solution vector (u∗,L) together with successful solver termination. The tolerance criterion ensured that successive updates to the solution were sufficiently small, while the solver termination flag verified that a valid root of the nonlinear system had been obtained.

3) COMPUTATION DETAILS FOR COARE

The COARE version 3.0 (COAREv3, https://github.com/NOAA-PSL/COARE-algorithm, Fairall et al. 2003) bulk aerodynamic algorithm was utilized to estimate surface exchange fluxes from atmospheric and surface state variables. Unlike simpler bulk models, COARE-3 incorporates physical parameterizations derived from extensive field observations, allowing for the calculation of the momentum flux, sensible heat flux, and latent heat flux across a wide range of meteorological conditions. By iteratively solving for the stability parameters of the atmospheric surface layer, the algorithm accounts for the non-linear relationship between wind speed and surface roughness.

### *b. Offshore Data*

This study develops machine learning models based on three observational datasets containing ocean and atmospheric measurements as well as near-surface flux measurements of momentum and heat. The datasets included in this study are the FINO1 research platform (Sans Rodrigo 2011, Munoz-Esparza 2012), the CASPER-West FLIP research vessel (Ortiz-Suslow et al. 2019), and the Martha's Vineyard Coastal Observatory (MVCO) air-sea interaction tower (Austin et al. 2000, Chang and Kirincich 2025). Because we plan to combine data from all three sites into a single model, the atmospheric and ocean state variables included in this study consider those that are included in all three of these observational datasets. The common variables include wind speed, temperature, absolute or relative humidity, pressure, sensible heat flux, friction velocity (or momentum flux), sea surface temperature (SST), significant wave height, and wave phase speed. Quality-control (QC) measures taken in each dataset include removing erroneously high or low values as well as removing any timestamps for which all required variables are not present.

#### 1) FINO1 PLATFORM

The FINO1 platform is located in the North Sea roughly 45 km north of Borkum, Germany. The observational period of focus for this study are for the years 2006 and 2011. The meteorological mast on the platform measures atmospheric variables up to roughly 90 m above sea level including three levels (40.0 m, 60.0 m, and 80.0 m) with high-rate sonic anemometer measurement as well as ocean state variables. In this study, we include turbulent flux data from the lowest sonic anemometer level of 40.0 m. Data are 10-min averaged at hourly intervals (Munoz-Esparza 2012). The FINO1 dataset includes 5,177 samples after QC measures are taken.

### 2) CASPER-West FLIP

The CASPER-West field experiment observed atmospheric and ocean state variables using a network of buoys and other platforms off the coast of California in October of 2017. All meteorological and ocean state variables included in this study are from the FLIP research platform. Atmospheric variables from FLIP are from the air-sea interaction (ASI) mast which includes flux measurements at seven vertical levels. Data used in this study are from the lowest platform level of 4.69 m and are provided as 10-minute averages at 10-minute intervals. The resulting number of samples after QC measures are taken is 2,952.

### 3) MVCO ASIT

The Martha's Vineyard Coastal Observatory (MVCO) is located in Edgartown, Massachusetts and has recorded observations in the Atlantic just south of Martha's Vineyard from 2001 through 2024 (and ongoing as of the writing of this paper). The data used for this study come from the MVCO Air-Sea Interaction Tower (ASIT) from 2006 to 2016. Within this 10-year span, there are several months of missing or excluded data from the QC process. The data used in this study considers meteorological variables from the lowest level of the ASIT flux measurements of 18.4 m and is 20-minute averaged data at 20-minute intervals. The resulting number of samples after QC measures are taken is 53,278, an order of magnitude more than was available for the other two datasets.

Note that the kinematic sensible heat flux observations for MVCO are provided with two decimals of precision. When converting these values to sensible heat flux (by the multiplication of $\rho C_p$; $1216.0\ Wm^2 / Kms^{-1}$ this results in further loss of granularity within the data.

### 4) Data Splitting

To ensure robust model evaluation, data at each site were partitioned using k-fold cross-validation in combination with an independent holdout dataset reserved for final testing. The number of samples used for cross-validation was 46,137 at MVCO, 4,453 at FINO, and 2,526 at CASPER. Independent holdout datasets consisted of 7,142 samples at MVCO (~13%), 726 samples at FINO (~14%), and 427 samples at CASPER (~14%). The holdout datasets were constructed by defining a period of interest for simulation for each case, finding the days within these periods in which data recovery is at or above 90%, and then selecting roughly 15% of these "high recovery" days spread across the period of interest. Five-fold cross-validation was applied to the larger MVCO dataset, while 10-fold cross-validation was used for the smaller FINO and CASPER datasets. Within each fold, data were divided into training and validation subsets. The validation data were used for model evaluation during hyperparameter optimization and, for neural network models, to implement early stopping based on validation loss.

This approach ensured that the majority of available data (~85–87%) was used for model development through cross-validation, while preserving an independent holdout dataset for unbiased evaluation of final model performance.

### *c. Methods: ML Models and Hyperparameter Optimization*

Model predictor variables were measured or derived from physically relevant quantities, including wind speed, relative humidity, pressure, sea surface temperature, and wave characteristics, and pressure. The derived variables included the Bulk Richardson Number, $Ri_B$ used as stability regime indicator, and vertical gradients (eg. $\frac{dU}{dz}, \frac{dT}{dz}, \frac{dq_v}{dz}$). Gradients were used as predictors rather than direct observations to normalize for differences in sensor height across sites. The gradient-based representation provides a height-normalized description of near surface atmospheric structure, consistent with surface-layer similarity theory, in which turbulent fluxes are governed by vertical gradients. Note that the use of gradients was also found preferable for applications in prior work (Munoz-Esparza et al. 2022).

Two machine learning approaches were evaluated: Random Forest models implemented using *Scikit-learn* and a custom dense neural network implemented in *TensorFlow*.

#### 1) Random Forest

Random Forest models were implemented as ensembles of decision trees trained on bootstrapped subsets of the data. At each split, a random subset of predictors was considered to promote model diversity and reduce overfitting. Key hyperparameters included the number of trees, maximum tree depth, number of features considered at each split, and the minimum number of samples required for splitting and leaf nodes. Tables A1 and A2 in the Appendix list the hyperparameters.

2) NEURAL NETWORK

The neural network consisted of a configurable number of fully connected hidden layers with nonlinear activation functions and a final linear output layer for regression. Model parameters were optimized using the Adam optimizer, and early stopping based on validation loss was employed to prevent overfitting. Input predictors and target variables were normalized prior to training. Sample weights were incorporated into the loss function when appropriate to account for inter-site data imbalance (see Section 2b4).

Hyperparameters included the number of hidden layers, number of neurons per layer, activation function (e.g., ReLU, Swish), learning rate (log-uniform range 10−4,10−2), batch size, and L2 regularization strength. Tables A1 and A2 in the Appendix list the optimal hyperparameters.

3) HYPERPARAMETER OPTIMIZATION AND PREDICTOR SELECTION

Hyperparameters for both model types were optimized using Bayesian optimization implemented with Optuna (Akiba et al. 2019). This approach iteratively refines the parameter search space by leveraging information from previous trials, providing improved efficiency compared to exhaustive grid search. For each site, flux type, and model type, 400 optimization trials were conducted using the full set of candidate predictors. This number of trials was selected to balance computational cost with adequate exploration of the hyperparameter space, as each trial required full model training and evaluation across all cross-validation folds. While additional trials may further refine the solution, this configuration was sufficient to identify well-performing regions of the parameter space.

Hyperparameter optimization and predictor selection were performed using a two-stage process. An initial Bayesian optimization was conducted using the full set of predictors, followed by permutation-based importance testing to identify the most influential variables.

When the resulting predictor set differed from the full set, a second stage of Bayesian optimization was performed using the reduced predictor set, followed by a final importance analysis. This iterative procedure enabled joint refinement of model structure and feature space, ensuring that hyperparameters were consistent with the final set of physically-relevant predictors.

Across all datasets and flux types, this process consistently identified a core set of physically meaningful predictors, including wind speed, vertical gradients of velocity and temperature, ocean state, and thermodynamic variables. Random Forest models more frequently required a second optimization stage, indicating greater sensitivity to predictor selection, whereas NN models generally retained similar predictor sets after the initial optimization.

For the RF, although the optimal hyperparameter search frequently selected unconstrained maximum tree depth (i.e., max_depth=None), further analysis of the trained ensembles revealed that the effective tree depth remained finite and varied systematically across datasets.

Specifically, deeper trees were observed for the MVCO and combined datasets (mean depths of approximately 37–42), while shallower trees were sufficient for the smaller FINO and CASPER datasets (mean depths of approximately 25–26). This indicates that tree growth was primarily governed by data characteristics rather than imposed hyperparameter limits.

An exception was observed for the CASPER momentum flux model, for which the optimal configuration included a constrained maximum depth of 20, with all trees reaching this limit. This suggests that limiting tree depth improved generalization performance in this data-limited regime by reducing overfitting.

For the neural network models, early stopping was employed with a patience parameter of 25 to prevent overfitting. Although L2 regularization strength was included as a tunable hyperparameter, it was ultimately set to zero in the final models, as this configuration improved the representation of extreme values in the target distributions. Final predictor sets and model configurations for all experiments are summarized in the tables in Appendix A.

*d. Combined Site Data*

In addition to developing site-specific datasets and models, a combined dataset was constructed to train a unified model across all sites, enabling the model to learn from a broader range of environmental conditions. Because the sites contributed unequal numbers of observations, site-based sample weighting was applied during model training to prevent the most heavily sampled site from dominating the fitted model. The weighting procedure was based on median-frequency balancing, in which weights are inversely proportional to class frequency. (Badrinarayanan et al., 2017)

Sample weights were calculated separately for each cross-validation fold using only the corresponding training data. For each site s, the number of training samples $N_s$ was first determined, and the median site sample count, $N_{med}$, was used as a reference value. Each sample from site s was initially assigned a weight, $w_s = \frac{N_{med}}{N_s}$, where $w_s$ is the initial weight assigned to all samples originating from site s. This formulation assigned weights near unity to FINO, the median-sized dataset; weights greater than one to FLIP, the least-represented site; and weights less than one to MVCO, the most-represented site.

The resulting weights were normalized to have a mean value of approximately 1.0 before being clipped to the range [0.25,4.0]. Mean normalization maintained a consistent overall weight scale during model fitting, while clipping limits extreme differences in the influence of individual samples. The clipping bounds were selected as a practical regularization measure to ensure that observations from the most heavily sampled site were not excessively downweighted and observations from the least represented site were not disproportionately emphasized.

The training-derived site weights were also applied to the corresponding validation data within each fold so that neural-network early stopping and model selection were based on the same balanced objective used during training. For final evaluation, the trained combined model was applied separately to the unweighted holdout dataset from each site, and performance metrics were calculated independently by site. This prevented MVCO from dominating model development while allowing site-specific holdout performance to be assessed directly.

Both the RF and NN models incorporated these weights during training. In the NN, the weights scale the contribution of each sample to the loss function, thereby modulating gradient updates during backpropagation. In the RF, sample weights affected the weighted impurity reduction used to evaluate candidate tree splits, increasing the influence of observations from less-represented sites.

The resulting sample weights were approximately 0.25 for MVCO, 1.05 for FINO, and 1.85 for FLIP. Thus, an individual MVCO observation contributed approximately one-quarter as much as a unit-weight observation, whereas observations from FLIP received greater influence during model fitting. Because MVCO contained substantially more observations, however, it continued to make a meaningful aggregate contribution to training. Overall, the weighting procedure reduced inter-site sampling imbalance while retaining all available observations and applying a consistent training strategy across both model types.

*e. Metrics for Assessment*

The primary metrics used in this analysis include the usual calculation methods for root mean squared error (RMSE), mean absolute error (MAE), and Pearson correlation coefficient (R). In addition, we assess using the Wasserstein Distance (also known as the Earth Movers Distance), which measures the dissimilarity between two probability distribution functions (smaller values denote better results) (Givens et al. 1984).

## 3. Results

The RF and NN models were developed, optimized, and tested for each of the individual sites (section 3a). As described above, different amounts of training and testing data were available for each of the three sites studied. Thus, we applied the models built for the MVCO site, which had the largest number of samples, to the other two sites to assess whether the model is likely to transfer (section 3b). Then data were combined from all sites to produce ML models to test on each site (section 3c). Finally, we assess variable importance and whether adding the derived variable, $Ri_B$ adds value (section 3c).

*a. Site Specific Results*

As a first test of the method, we build ML models for each of the three sites and test each of the models at the site for which it was configured and optimized.

1) Site Specific Momentum Flux Estimates

The results of momentum flux prediction experiments are indicated in Figure 1 and Table 1. The machine learning models shine in terms of MAE, RMSE, and Pearson correlation at the MVCO and FINO sites, with the exception of the COARE-3 model showing the lowest MAE at FINO. At the CASPER site, however, the COARE–3 algorithm is optimal. We note that this site has the least amount of data for training the ML algorithms. The results for the Wasserstein distance are more variable, with MOST being best for MVCO, NN for FINO, and COARE-3 for CASPER. Whether the NN or RF performed best varied with the site and the metric, but with the RF showing superiority for more metrics.

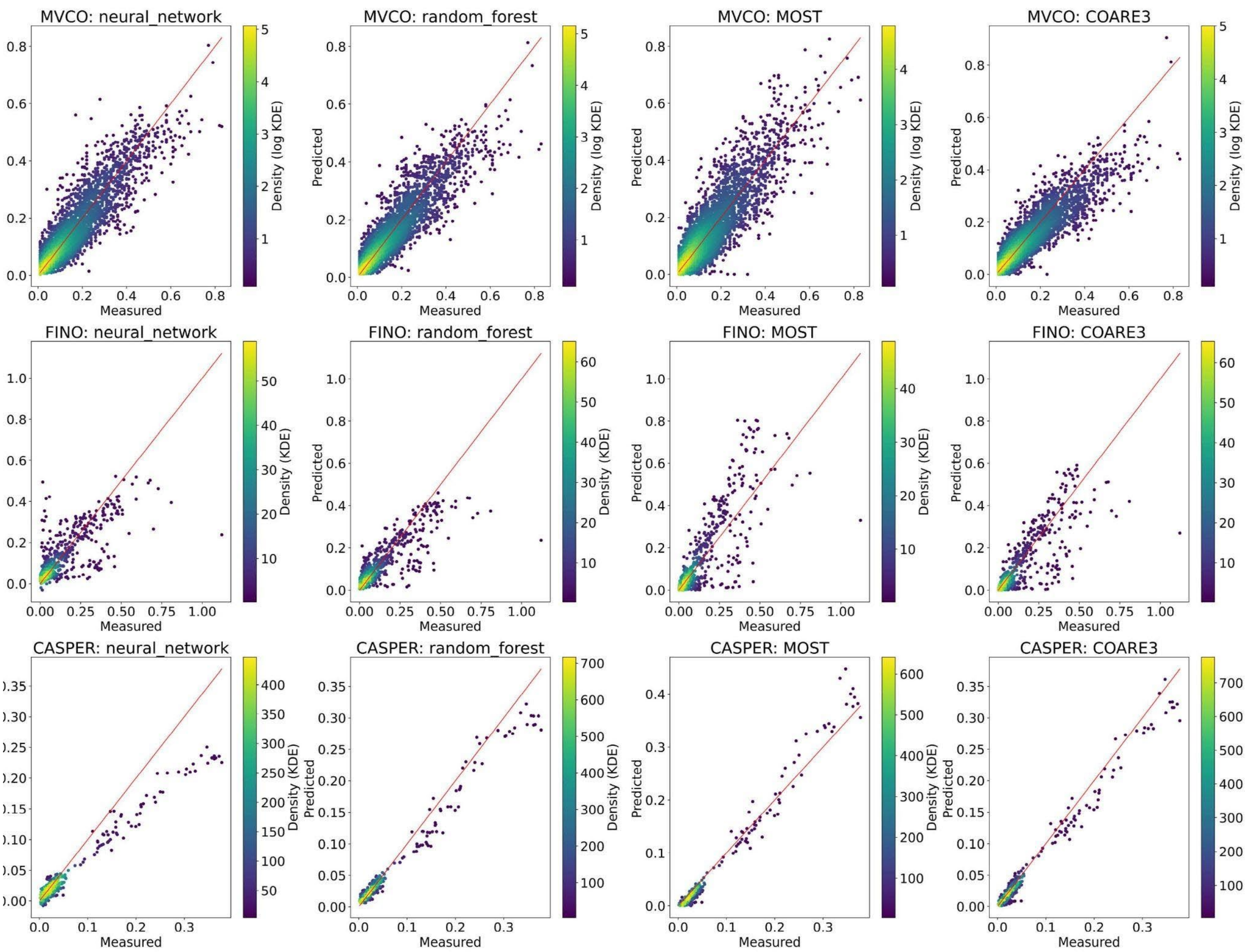

Figure 1. Scatter plots of observed (abscissa) versus modeled (ordinate) momentum flux. The first column of each row shows results for the neural network ML model, the second column for the random forest, the third column for MOST, and the fourth column for COARE-3. The first row is the model built and applied for the MVCO site, the second row for the FINO site, and the third row for the CASPER data.

Table 1. Comparison of the various methods for estimating momentum flux for each of the three sites. The best value of each metric at each site is highlighted in bold.

| Site | Method | RMSE | MAE | Pearson Correlation | Wasserstein Distance |
|---|---|---|---|---|---|
| **MVCO** | Neural Net | 4.39e-2 | 2.88e-2 | 0.91 | 5.03e-3 |
| | Random Forest | **4.24e-2** | **2.74e-2** | **0.92** | 6.84e-3 |
| | MOST | 5.19e-2 | 3.55e-2 | 0.89 | **4.64e-3** |
| | COARE-3 | 4.52e-2 | 2.94e-2 | 0.91 | 1.05e-2 |
| **FINO** | Neural Net | 9.01e-2 | 4.93e-2 | 0.77 | **1.69e-2** |
| | Random Forest | **7.79e-2** | 4.35e-2 | **0.85** | 2.40e-2 |
| | MOST | 9.78e-2 | 5.70e-2 | 0.83 | 2.14e-2 |
| | COARE-3 | 8.18e-2 | **4.32e-2** | 0.83 | 1.80e-2 |
| **CASPER WEST** | Neural Net | 2.83e-2 | 1.53e-2 | 0.98 | 1.11e-2 |
| | Random Forest | 1.49e-2 | 8.86e-3 | **0.99** | 6.11e-3 |
| | MOST | 1.50e-2 | 1.05e-2 | **0.99** | 9.50e-3 |
| | COARE-3 | **1.24e-2** | **7.58e-3** | **0.99** | **5.06e-3** |

2) SITE SPECIFIC HEAT FLUX ESTIMATES

For heat flux (Figure 2, Table 2), the machine learning models outperform physics-based parameterizations in the offshore surface layer across all sites, except that the correlation was highest for MOST at the CASPER WEST site. Both NN and RF models achieved substantially lower RMSE, MAE, and Wasserstein Distance compared to MOST and generally outperformed or matched COARE.

The performance of the NN and the RF models is comparable for MVCO and FINO, the two largest datasets. For the CASPER site, the RF outperforms the NN. Thus both ML approaches were effective but neither consistently dominated all heat flux experiments when assessed for the site for which they were trained.

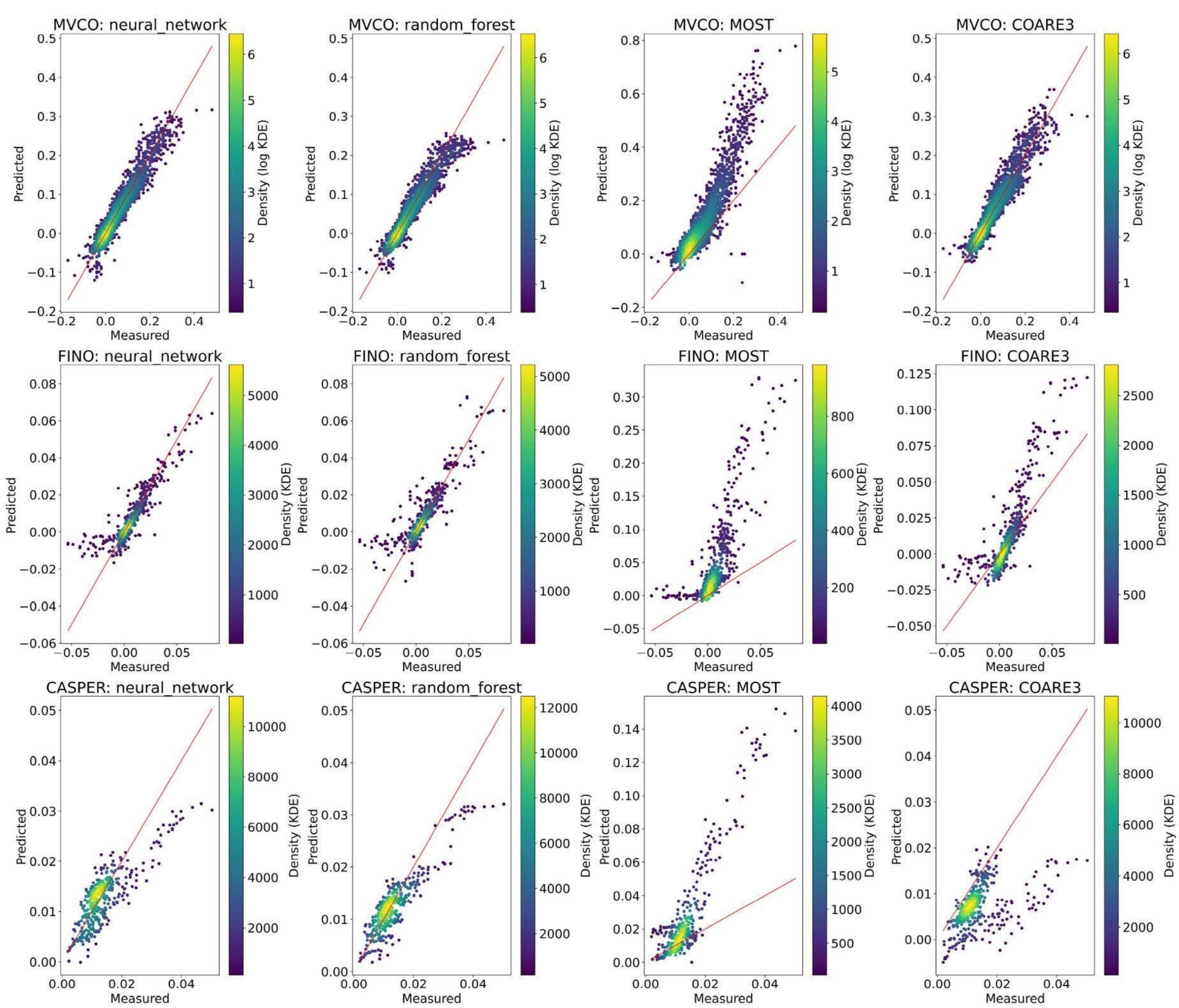

Figure 2. As for Fig. 1 but for heat flux.

Table 2. Comparison of the various methods for estimating heat flux for each of the three sites. The best value of each metric at each site is highlighted in bold.

| Site | Method | RMSE | MAE | Pearson Correlation | Wasserstein Distance |
|---|---|---|---|---|---|
| **MVCO** | Neural Net | **1.68e-2** | 1.20e-2 | **0.96** | **3.65e-3** |
| | Random Forest | 1.74e-2 | **1.18e-2** | **0.96** | 4.35e-3 |
| | MOST | 6.16e-2 | 3.72e-2 | 0.91 | 3.53e-2 |
| | COARE-3 | 1.75e-2 | 1.21e-2 | 0.95 | 3.51e-2 |
| **FINO** | Neural Net | **7.00e-3** | **3.87e-3** | **0.88** | 1.67e-3 |
| | Random Forest | 7.55e-3 | 4.17e-3 | 0.86 | **1.53e-3** |
| | MOST | 6.08e-2 | 3.42e-2 | 0.79 | 3.40e-2 |
| | COARE-3 | 1.48e-2 | 9.01e-3 | 0.84 | 7.13e-3 |
| **CASPER WEST** | Neural Net | 4.22e-3 | 3.16e-3 | 0.83 | 1.51e-3 |
| | Random Forest | **3.46e-3** | **2.61e-3** | 0.90 | **1.06e-3** |
| | MOST | 2.61e-2 | 1.42e-2 | **0.92** | 1.30e-2 |
| | COARE-3 | 9.05e-3 | 6.46e-3 | 0.48 | 6.28e-3 |

*b. MVCO Application to Other Sites*

The question arises as to whether an ML surface layer model built at one site would perform well when applied to a different site. To answer this question, we tested applying the ML models built for the MVCO site, the site with the most training/testing data, could be applied to the other two sites. Figures 3 and 4 show the scatter plots for this experiment for momentum flux and heat flux, respectively.

For momentum flux, Table 3 shows that in all cases, the metrics are degraded from when using the ML model trained for the specific site (cf. Table 1). The RF performs better than the NN for RMSE, MAE, and Wasserstein distance, but the NN performs better in terms of correlation at the CASPER West site, although Fig. 3 demonstrates that the measured values are nearly constant and anomolously small, calling into question the relationship. When compared to COARE-3, however, the ML models do not perform as well when applied to a site they were not trained on.

The results for heat flux are rather interesting. As seen in Fig. 4 and Table 4, at FINO, although the MVCO ML models did not perform as well as those trained for that site (cf. Table 2), both the RF performed better than MOST or COARE-3 in terms of RMSE and MAE, but not for correlation. For the CASPER site, the results for COARE-3 continue to outperform any other method, except that MOST and NN both showed a higher correlation.

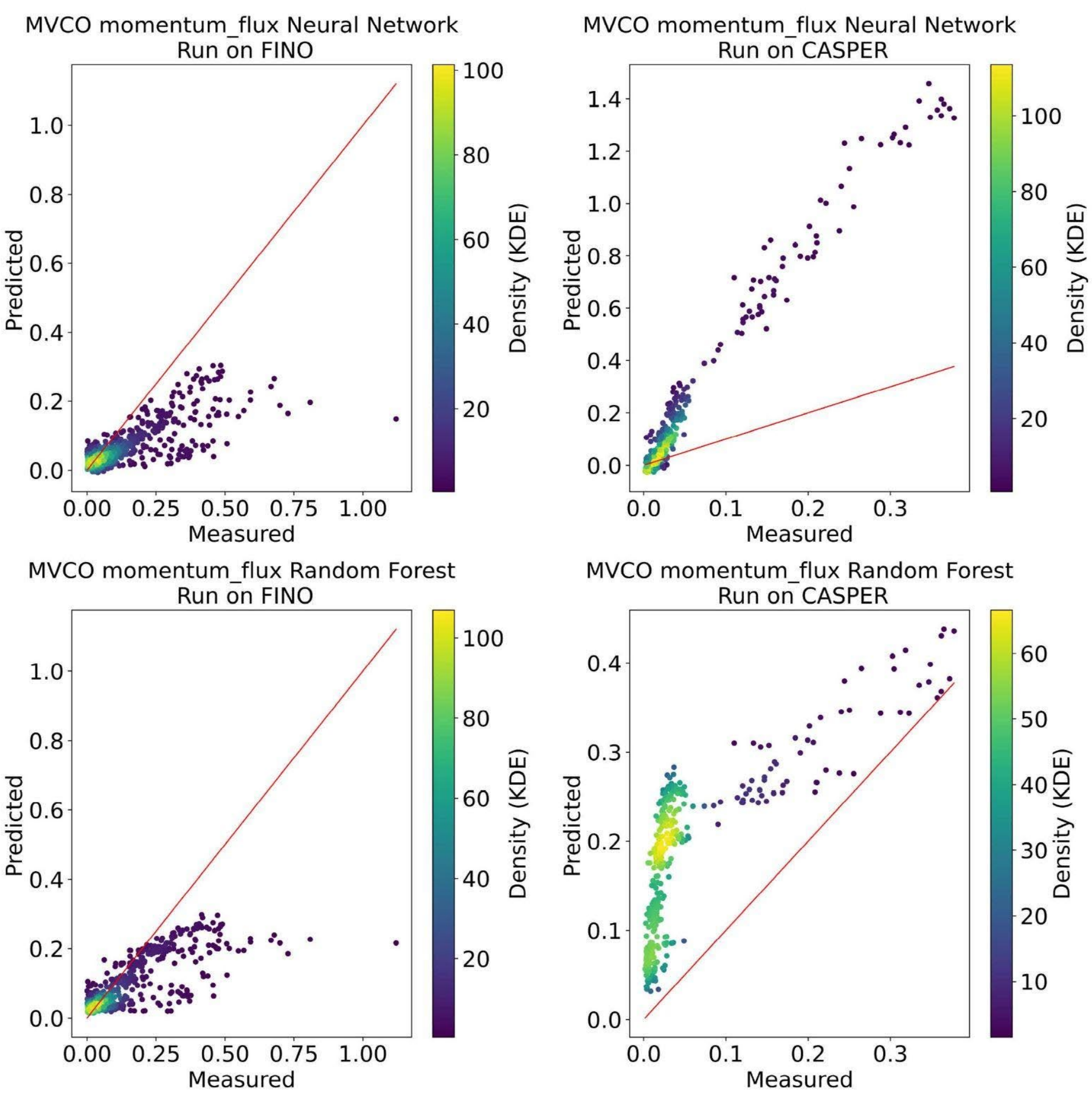

Figure 3. Scatter plots of the momentum flux computed with the MVCO ML models applied to the FINO (left columns) and CASPER WEST (right columns) sites. The top row shows the results for the NN and the bottom row for the RF.

Table 3. Comparison of the momentum flux computed with MVCO model applied at the FINO and CASPER-WEST sites.

| Site | Method | RMSE | MAE | Pearson Correlation | Wasserstein Distance |
|---|---|---|---|---|---|
| **FINO** | Neural Net | 1.11e-1 | 6.64e-2 | 0.78 | 0.06 |
| | Random Forest | 1.03e-1 | 5.43e-2 | 0.80 | 0.05 |
| | MOST | 9.78e-2 | 5.70e-2 | **0.83** | **0.02** |
| | COARE-3 | **8.18e-2** | **4.32e-2** | **0.83** | **0.02** |
| **CASPER WEST** | Neural Net | 2.82e-1 | 1.47e-1 | 0.98 | 0.15 |
| | Random Forest | 1.43e-1 | 1.30e-1 | 0.73 | 0.13 |
| | MOST | 1.50e-2 | 1.05e-2 | **0.99** | 0.09 |
| | COARE-3 | **1.24e-2** | **7.58e-3** | **0.99** | **0.05** |

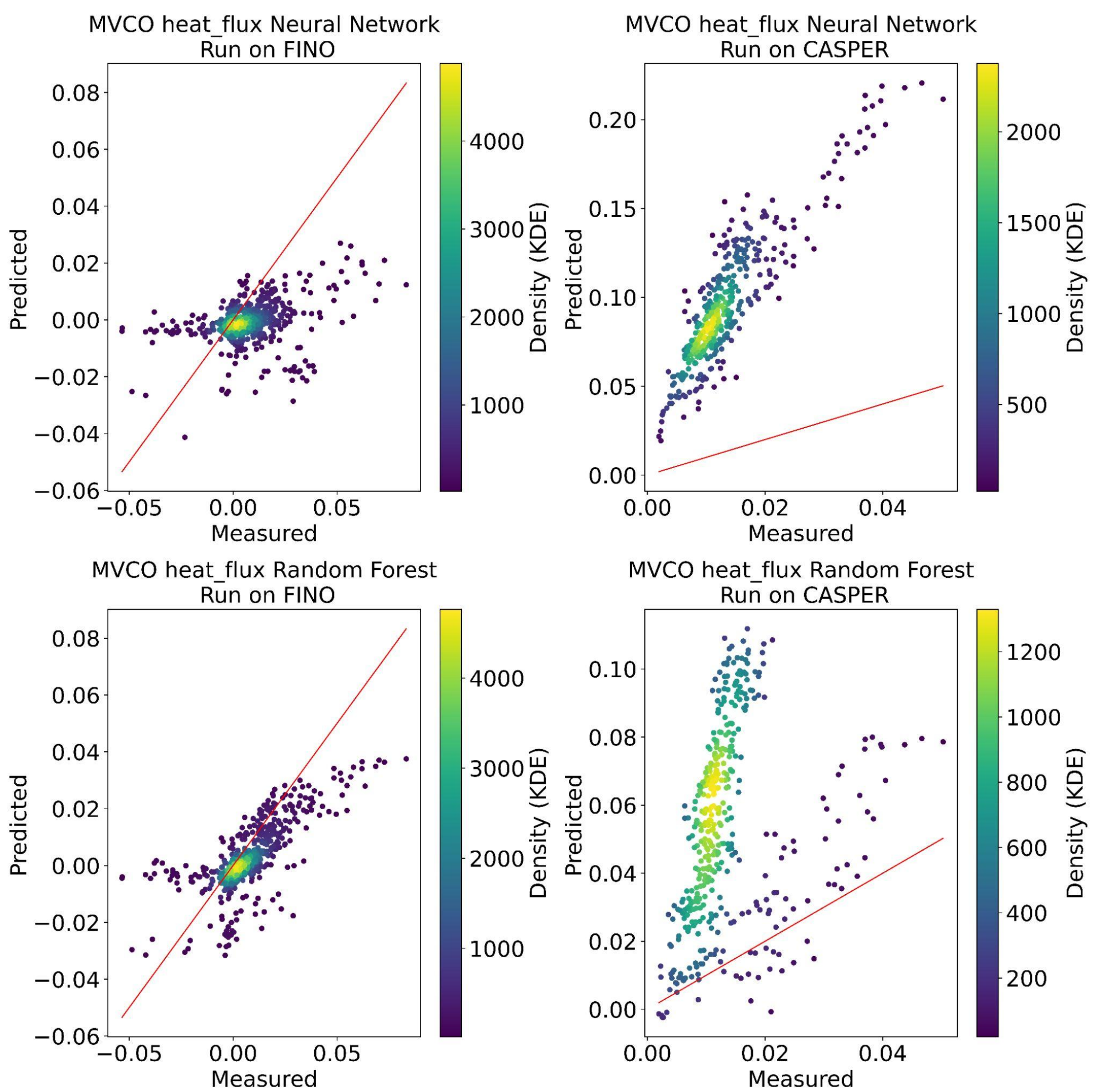

Figure 4. As for Fig. 3, but for heat flux.

Table 4. Comparison of the heat flux computed with MVCO model applied at the FINO and CASPER-WEST sites.

| Site | Method | RMSE | MAE | Pearson Correlation | Wasserstein Distance |
|---|---|---|---|---|---|
| **FINO** | Neural Net | 1.58e-2 | 1.07e-2 | 0.34 | 0.01 |
| | Random Forest | **1.10e-2** | **7.48e-3** | 0.73 | **0.00** |
| | MOST | 6.08e-2 | 3.42e-2 | 0.79 | 0.03 |
| | COARE-3 | 1.48e-2 | 9.01e-3 | **0.84** | 0.07 |
| **CASPER WEST** | Neural Net | 8.73e-2 | 8.24e-2 | 0.89 | 0.08 |
| | Random Forest | 4.84e-2 | 4.11e-2 | 0.26 | 0.04 |
| | MOST | 2.61e-2 | 1.42e-2 | **0.92** | **0.01** |
| | COARE-3 | **9.05e-3** | **6.46e-3** | 0.48 | **0.01** |

3) COMBINED DATA FLUX ESTIMATES

Although the MVCO ML models transferred well to the FINO site, we saw limited success at applying the MVCO ML models to the CASPER site, which has disparate characteristics from the two Atlantic sites. Thus, we seek to answer the question as to whether combining data from all three sites would produce a better ML model than using data from a single model.

1) MOMENTUM FLUX

Table 5 shows the results for the combined data models applied to each of the three sites. The results for the MVCO site are quite similar to those for the model built only on data from that site, with the ML models, particularly the RF, outshining the MOST and COARE-3 models for all metrics except Wasserstein distance, for which MOST won out. For the FINO site, the RF outperformed the other methods, except for NN showing the best Wasserstein distance, but note that the COARE-3 algorithm's performance was closely matched. When compared to the ML models trained only on FINO data (cf. Table 1), the performance of the NN improved slightly, but the RF did not. Finally, for the CASPER WEST site, although the ML algorithms trained on data from all sites performed better than from that single site, the COARE-3 algorithm outperformed them. In general, all the models, ML and semi-empirical, perform well estimating the surface momentum flux.

Table 5. Comparison of the various methods for estimating momentum flux for each of the three sites using data combined from all sites.

| Site | Method | RMSE | MAE | Pearson Correlation | Wasserstein Distance |
|---|---|---|---|---|---|
| **MVCO** | Neural Net - Combined | 4.55e-2 | 2.98e-2 | 0.91 | 5.71e-3 |
| | Random Forest - Combined | **4.24e-2** | **2.75e-2** | **0.92** | 6.65e-3 |
| | MOST | 5.19e-2 | 3.55e-2 | 0.89 | **4.64e-3** |
| | COARE-3 | 4.52e-2 | 2.94e-2 | 0.91 | 1.05e-2 |
| **FINO** | Neural Net- Combined | 8.65e-2 | 4.67e-2 | 0.80 | **1.79e-2** |
| | Random Forest- Combined | **8.12e-2** | **4.31e-2** | **0.83** | 2.30e-2 |
| | MOST | 9.78e-2 | 5.70e-2 | **0.83** | 2.14e-2 |
| | COARE-3 | 8.18e-2 | 4.32e-2 | **0.83** | 1.80e-2 |
| **CASPER WEST** | Neural Net- Combined | 2.42e-2 | 1.52e-2 | 0.97 | 1.18e-2 |
| | Random Forest - Combined | 1.46e-2 | 9.23e-3 | **0.99** | 7.31e-3 |
| | MOST | 1.50e-2 | 1.05e-2 | **0.99** | 9.50e-3 |
| | COARE-3 | **1.24e-2** | **7.58e-3** | **0.99** | **5.06e-3** |

For the NN, combined training improved the pointwise metrics at FINO (RMSE, MAE, Pearson Correlation) and yielded some gains for CASPER (most notably the RMSE).

Performance was degraded across all metrics for MVCO when compared to the model trained only at that site.

In contrast, RF models exhibited mixed behavior with minor improvements in some metrics (e.g., Wasserstein distance at MVCO, MAE at FINO and CASPER) but no overall consistent benefit.

This suggests that momentum flux can often be predicted using an ML model from local conditions alone even when using a low-volume training data set, with limited and model dependent gains from incorporating cross-site training data. This finding is consistent with the relatively strong baseline performance of both ML and physics-based models for this variable.

2) Heat Flux Estimates

Table 6 shows results for combining all data for ML training for heat flux. NN models trained on combined data showed a clear benefit for the site with the smallest dataset (CASPER, cf. Table 2), where performance improved across all metrics. There was a slight degradation for the NN estimates of heat flux at FINO, although the changes were modest, while changes were small and mixed for MVCO, depending on the metric assessed.

In contrast, RF models generally exhibited degraded performance under combined training for the FINO and CASPER sites, but with very little change for MVCO. This suggests that combined data training benefits NNs in data limited areas, whereas RF models are more sensitive to cross-site heterogeneity.

In general, when data are combined from all sites, the NN was the best model for both metrics and both NN and RF beat the MOST and COARE-3 models.

Table 6. Comparison of the various methods for estimating heat flux for each of the three sites using data combined from all sites.

| Site | Method | RMSE | MAE | Pearson Correlation | Wasserstein Distance |
|---|---|---|---|---|---|
| **MVCO** | Neural Net-Combined | **1.65e-2** | **1.14e-2** | **0.96** | 4.74e-3 |
| | Random Forest-Combined | 1.74e-2 | 1.17e-2 | **0.96** | **4.35e-3** |
| | MOST | 6.16e-2 | 3.72e-2 | 0.91 | 3.53e-2 |
| | COARE-3 | 1.75e-2 | 1.21e-2 | 0.95 | 3.51e-2 |
| **FINO** | Neural Net-Combined | **7.64e-3** | **4.33e-3** | **0.86** | **1.82e-3** |
| | Random Forest-Combined | 7.70e-3 | 4.61e-3 | **0.86** | 2.63e-3 |
| | MOST | 6.08e-2 | 3.42e-2 | 0.79 | 3.40e-2 |
| | COARE-3 | 1.48e-2 | 9.01e-3 | 0.84 | 7.13e-3 |
| **CASPER WEST** | Neural Net-Combined | **2.79e-3** | **2.20e-3** | **0.93** | **7.28e-4** |
| | Random Forest-Combined | 6.40e-3 | 4.28e-3 | 0.70 | 2.45e-3 |
| | MOST | 2.61e-2 | 1.42e-2 | 0.92 | 1.30e-2 |
| | COARE-3 | 9.05e-3 | 6.46e-3 | 0.48 | 6.28e-3 |

*d. Relative importance of variables and derived variables*

Analyzing the relative importance of the variables used provides insight into the physical interpretation of the ML models. In addition, we chose variables with physical insight and included at least one derived variable, Bulk Richardson number ( $Ri_B$ ), that indicates boundary layer stability. We wish to assess these issues as a way of extending the explainability of the ML models.

1) VARIABLE IMPORTANCE

We analyze the variable importance results for the ML algorithms built on the combined data in Fig. 5. The momentum flux results show that for both the NN and RF, wind speed is most important by a large margin. For the NN, this is followed by the temperature gradient, wave height, wind speed gradient, relative humidity, pressure, wave phase speed, SST, and moisture gradient. The $Ri_B$ is the least important variable in this case. For the RF the wind speed gradient and temperature gradient were the next most important variables, followed by moisture gradient, $Ri_B$, SST, and wave height, with pressure and relative humidity being of least importance.

For heat flux, both the NN and RF ranked the temperature gradient as the most important variable by a large margin, followed by wind speed, wind speed gradient, and moisture gradient. The NN then ranked wave height and relative humidity next, while the RF used $Ri_B$ and SST. The remaining variables are fairly unimportant.

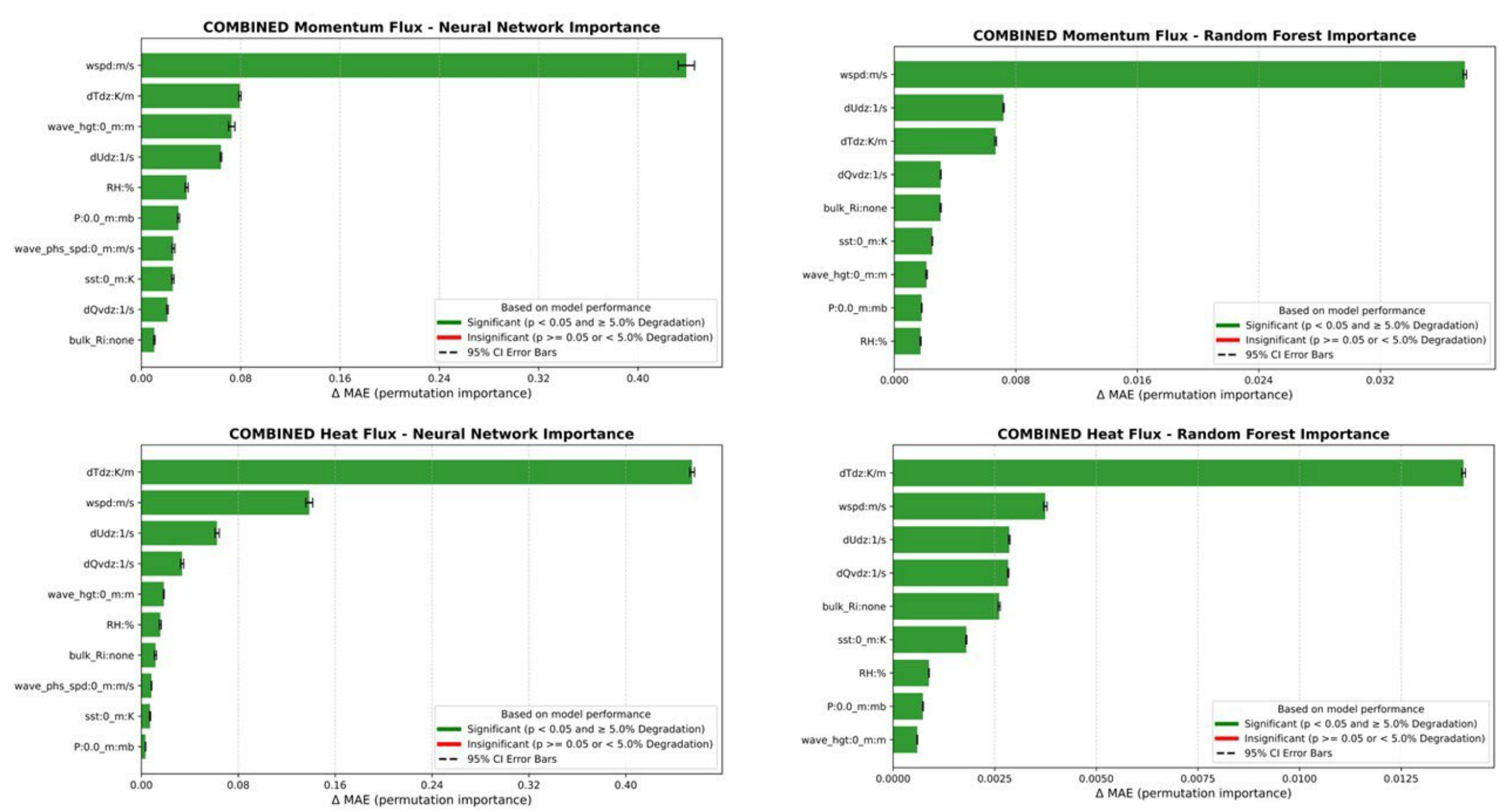

Figure 5. Bar plots of variable importance for momentum flux (top row) and heat flux (bottom row). The left column is for NN while the right column represents RF results.

#### 2) Does Bulk Richardson number provide value?

Because Bulk Richardson number was derived from available input variables with an intent of distilling information on the stability of the lower atmosphere to the ML models, we analyze whether or not the models improve or degrade when it is left out. We already noted in 3d1 that $Ri_B$ was of fairly low importance for the combined models.

Table 7 compares the neural network and random forest models on the RMSE, MAE, and Pearson correlation coefficient metrics with and without $Ri_B$ as a derived variable for training to estimate momentum flux using data from all sites combined and applied to each of the three sites. In general, differences are small and inconsistent. For the MVCO site using a NN, the RMSE and MAE are smaller when the $Ri_B$ is included, but the opposite is true for the RF. There is no difference in Pearson correlation. For the FINO site, both models produced slightly better results without $Ri_B$ where there were differences. At the CASPER site, both models tended to be slightly better without using $Ri_B$, except for a slight improvement in the RF RMSE.

Table 7. Comparison of using Bulk Richardson number as a derived variable for estimating momentum flux using data from all sites and evaluated for each of the three sites. The best value of each metric at each site is highlighted in bold.

| Site | Method | RMSE | | MAE | | Pearson Correlation | |
|---|---|---|---|---|---|---|---|
| | | With Ri_B | Without Ri_B | With Ri_B | Without Ri_B | With Ri_B | Without Ri_B |
| **MVCO** | Neural Net | **4.39e-2** | 4.45e-2 | **2.88e-2** | 2.93e-2 | 0.91 | 0.91 |
| | Random Forest | 4.24e-2 | **4.23e-2** | 2.74e-2 | **2.73e-2** | 0.92 | 0.92 |
| **FINO** | Neural Net | 9.01e-2 | **7.99e-2** | 4.93e-2 | **4.41e-2** | 0.77 | **0.82** |
| | Random Forest | 7.79e-2 | **7.73e-2** | 4.35e-2 | 4.35e-2 | 0.85 | 0.85 |
| **CASPER WEST** | Neural Net | 2.83e-2 | **1.71e-2** | 1.53e-2 | **1.12e-2** | 0.98 | **0.99** |
| | Random Forest | **1.49e-2** | 1.50e-2 | 8.86e-3 | **8.85e-3** | 0.99 | 0.99 |

Table 8 similarly presents comparison results for heat flux. Again, differences tend to be relatively small and there is a mixed message. The random forest is somewhat better for MVCO when using $Ri_B$, but the neural network is not. Results are uneven for the FINO site, but for the CASPER site, both models are improved when including $Ri_B$.

In general, it appears that the models have used the raw information, including the gradients of wind speed and moisture, to discover the importance of stability for estimating the surface fluxes of momentum and heat, but there are cases where explicitly including $Ri_B$ is helpful.

Table 8. Comparison of using Bulk Richardson number as a derived variable for estimating heat flux using data from all sites and evaluated for each of the three sites. The best value of each metric at each site is highlighted in bold.

| **Site** | **Method** | **RMSE** | | **MAE** | | **Pearson Correlation** | |
|---|---|---|---|---|---|---|---|
| | | **With Ri_B** | **Without Ri_B** | **With Ri_B** | **Without Ri_B** | **With Ri_B** | **Without Ri_B** |
| **MVCO** | Neural Net | 1.68e-2 | **1.62e-2** | 1.20e-2 | **1.15e-2** | 0.96 | 0.96 |
| | Random Forest | **1.74e-**2 | 1.85e-2 | **1.18e-2** | 1.21e-2 | **0.96** | 0.95 |
| **FINO** | Neural Net | 7.00e-3 | **6.99e-3** | **3.87e-3** | 3.91e-3 | 0.88 | 0.88 |
| | Random Forest | **7.55e-3** | 7.72e-3 | **4.17e-3** | 4.27e-3 | 0.86 | 0.86 |
| **CASPER WEST** | Neural Net | **4.22e-3** | 4.66e-3 | **3.16e-3** | 3.48e-3 | **0.83** | 0.78 |
| | Random Forest | 3**.46e-3** | 3.53e-3 | **2.61e-3** | 2.66e-3 | 0.90 | 0.90 |

## 4. Conclusions

We have used data from three disparate offshore sites to build both NN and RF models of surface fluxes of heat and momentum. We employ gradients of the wind and temperature variables to allow us to combine data from the three sites whose towers have differing measurement heights. The ML models built on the combined data generally outperformed those that were built for the individual sites, suggesting that such a model might be applicable for other offshore sites.

A clear distinction emerges between heat and momentum flux prediction, with important implications for improving numerical weather prediction in the offshore surface layer. Machine learning methods, particularly NNs, provide the greatest benefit for heat flux, where traditional physics-based parameterizations (e.g., MOST) perform poorly, while schemes specific to the offshore environment, such as COARE-3, are generally competitive but do not consistently match the ML performance if a sufficient amount of training data are available. In contrast, for momentum flux, both physics-based, semi-empirical models and ML models perform very well.

These results indicate that the potential gains from ML strongly depend on the underlying physical process. Heat flux, which depends on complex interactions among temperature gradients, humidity, stability, and surface exchange processes, is not fully captured by existing parameterizations. In this case, ML models—both NNs and RFs—provide substantial improvements in accuracy and in representing the full distribution of flux values. This suggests that ML-based parameterizations offer a promising pathway for improving heat flux representation in offshore NWP models.

In contrast, momentum flux is more directly related to wind speed and wind shear and is already well represented by existing physical parameterizations. As a result, machine learning provides only modest improvements, and in some cases physics-based models perform comparably. This indicates that momentum flux parameterizations are not substantially improved, and the incremental benefit of ML is smaller.

The impact of training strategy further supports this distinction. The use of combined multi-site data significantly improved NN performance for heat flux in data-limited regimes, demonstrating the ability of ML models to leverage diverse datasets to learn transferable relationships. However, combined training provided limited benefit for momentum flux and in some cases degraded RF performance, highlighting sensitivity to heterogeneous data distributions.

Overall, these findings suggest a hybrid pathway for improving offshore surface layer representation in numerical weather prediction: retain physics-based parameterizations where they are already effective (e.g., momentum flux), and deploy machine learning approaches where physical models are insufficient (e.g., heat flux), particularly in regimes characterized by limited observational data.

This study is limited to the three sites with available training data, which were mixed in terms of both quality and quantity of the data as well as heights of observations. We chose to formulate the models using gradients of wind speed and temperature, which allowed us to build the combined dataset. As the results of the combined data showed improvement in the ML models for all sites, this choice allowed that combination. The results indicate that given quality controlled data from disparate locations it is possible to develop ML models for estimation of heat fluxes substantially superior to currently available physics-based, semi-empirical models. This fact points to the need for more long-term offshore surface flux observations.

The aforementioned evaluation and comparison of machine learning outcomes were conducted using an offline methodology. Detailed analysis of the ML model performance within the context of numerical weather prediction implementation is provided in the companion study by Hawbecker et al. (2026).

*Acknowledgments.*

This material is based upon work supported by the NSF National Center for Atmospheric Research, which is a major facility sponsored by the U.S. National Science Foundation under Cooperative Agreement No. 1852977. The authors acknowledge support from the Observationally driven Resource Assessment with CoupLEd models (ORACLE) project under grant number 778383, sponsored by the U. S. Dept. of Energy and managed by Pacific Northwest National Laboratory (PNNL).

*Data Availability Statement.*

The MVCO dataset (DOI 10.26025/1912/29631) was downloaded from the MVCO Historical Data webpage: https://mvco.whoi.edu/data/data-history/ in March 2025. FINO and CASPER-WEST data were obtained by request from the authors referenced for those sites.

APPENDIX

## Hyperparameter settings for the ML Models

Table A1. Hyperparameter optimized settings for Heat Flux.

| Dataset | Flux | Model | 2nd Opt | Model Parameters | Final Predictors |
|---|---|---|---|---|---|
| MVCO | MF | NN | Yes | activation: relu<br>batch_size: 16<br>hidden_layers: 3<br>hidden_neurons: 128<br>lr: 0.00015<br>patience: 25<br>output_activation: linear | dUdz<br>wspd<br>dTdz<br>wave_hgt<br>dQvdz<br>RH<br>sst<br>wave_phs_spd<br>P |
| MVCO | MF | RF | Yes | max_depth: null<br>max_features: log2<br>min_samples_leaf: 2<br>min_samples_split: 2<br>n_estimators: 300 | dUdz<br>wspd<br>dTdz<br>Ri<br>RH<br>dQvdz<br>wave_hgt<br>sst |
| MVCO | HF | NN | No | activation: relu<br>batch_size: 32<br>hidden_layers: 4<br>hidden_neurons: 128<br>lr: 0.00179<br>patience: 25<br>output_activation: linear | dTdz<br>wspd<br>dUdz<br>dQvdz<br>wave_hgt<br>sst<br>RH<br>wave_phs_spd<br>Ri<br>P |
| MVCO | HF | RF | Yes | max_depth: null<br>max_features: sqrt<br>min_samples_leaf: 1<br>min_samples_split: 2<br>n_estimators: 300 | dTdz<br>dUdz<br>wspd<br>dQvdz<br>Ri<br>sst<br>RH<br>P<br>wave_hgt |

.

| Dataset | Flux | Model | 2nd Opt | Model Parameters | Final Predictors |
|---|---|---|---|---|---|
| FINO | MF | NN | No | activation: relu<br>batch_size: 32<br>hidden_layers: 4<br>hidden_neurons: 128<br>lr: 0.00106<br>patience: 25<br>output_activation: linear | dUdz<br>dTdz<br>wspd<br>wave_hgt<br>wave_phs_spd<br>sst<br>RH<br>P<br>dQvdz<br>Ri |
| FINO | MF | RF | No | max_depth: null<br>max_features: log2<br>min_samples_leaf: 2<br>min_samples_split: 2<br>n_estimators: 300 | dUdz<br>wspd<br>dTdz<br>wave_hgt<br>Ri<br>wave_phs_spd<br>P<br>RH<br>sst<br>dQvdz |
| FINO | HF | NN | No | activation: swish<br>batch_size: 64<br>hidden_layers: 3<br>hidden_neurons: 128<br>lr: 0.00014<br>patience: 25<br>output_activation: linear | dTdz<br>wave_hgt<br>dQvdz<br>dUdz<br>wspd<br>Ri<br>wave_phs_spd<br>RH<br>P<br>sst |
| FINO | HF | RF | Yes | max_depth: null<br>max_features: log2<br>min_samples_leaf: 2<br>min_samples_split: 2<br>n_estimators: 300 | dTdz<br>Ri<br>wspd<br>dUdz<br>wave_hgt<br>sst<br>dQvdz<br>RH<br>P |

Table A2. Hyperparameter optimized settings for Momentum Flux.

| Dataset | Flux | Model | 2nd Opt | Model Parameters | Final Predictors |
|---|---|---|---|---|---|
| Combined | MF | NN | No | activation: relu<br>batch_size: 16<br>hidden_layers: 3<br>hidden_neurons: 128<br>lr: 0.00022<br>l2_weight: 0.0<br>patience: 25<br>early_stop: True<br>output_activation: linear | wspd<br>dTdz<br>dUdz<br>RH<br>sst<br>wave_hgt<br>wave_phs_spd<br>dQvdz<br>P<br>Ri |
| Combined | MF | RF | Yes | max_depth: null<br>max_features: log2<br>min_samples_leaf: 1<br>min_samples_split: 2<br>n_estimators: 300 | wspd<br>dUdz<br>dTdz<br>dQvdz<br>Ri<br>sst<br>wave_hgt<br>RH<br>P |
| Combined | HF | NN | No | activation: swish<br>batch_size: 32<br>hidden_layers: 4<br>hidden_neurons: 64<br>lr: 0.00015<br>patience: 25<br>output_activation: linear | dTdz<br>wspd<br>dUdz<br>dQvdz<br>wave_hgt<br>RH<br>Ri<br>sst<br>wave_phs_spd<br>P |
| Combined | HF | RF | Yes | max_depth: null<br>max_features: log2<br>min_samples_leaf: 1<br>min_samples_split: 2<br>n_estimators: 300 | dTdz<br>wspd<br>dUdz<br>dQvdz<br>Ri<br>sst<br>RH<br>P<br>wave_hgt |